\documentclass{article}
\usepackage{graphicx} 
\PassOptionsToPackage{numbers, sort, compress}{natbib}
\usepackage[final]{neurips_2024}

\usepackage[utf8]{inputenc} 
\usepackage[T1]{fontenc}    
\usepackage{hyperref}       
\usepackage{url}            
\usepackage{booktabs}       
\usepackage{amsfonts}       
\usepackage{nicefrac}       
\usepackage{microtype}      
\usepackage{xcolor}         

\title{Evaluating Generative AI Systems \\ is a Social Science Measurement Challenge}
\author{%
\centerline{{
\small
\bf
Hanna Wallach$^1$ \ \ 
 Meera Desai$^2$ \ \ 
 Nicholas Pangakis$^1$ \ \ 
 A. Feder Cooper$^1$ \ \ 
 Angelina Wang$^3$
}} \\
\centerline{{
\small
\bf
Solon Barocas$^1$ \ \ 
 Alexandra Chouldechova$^1$\ \ 
Chad Atalla$^1$\ \  
Su Lin Blodgett$^1$
}} \\
\centerline{{
\small
\bf
Emily Corvi$^1$\ \ 
P. Alex Dow$^1$\ \ 
Jean Garcia-Gathright$^1$ \ \ 
Alexandra Olteanu$^1$
}}\\
\centerline{{
\small
\bf
Stefanie Reed$^1$ \ \ 
Emily Sheng$^1$\ \ 
Dan Vann$^1$ \ \ 
Jennifer Wortman Vaughan$^1$
}}\\
\centerline{{
\small
\bf
Matthew Vogel$^1$\ \ 
Hannah Washington$^1$ \ \ 
Abigail Z. Jacobs$^2$
}}\\
\centerline{{
\small
$^1$Microsoft Research \ \  $^2$University of Michigan \ \ 
$^3$Stanford University
}}
\\
\centerline{{
\small 
Corresponding email: \texttt{wallach@microsoft.com}
}}
}

\begin{document}

\maketitle

\begin{abstract}
    Across academia, industry, and government, there is an increasing awareness that the measurement tasks involved in evaluating generative AI (GenAI) systems are especially difficult. We argue that these measurement tasks are highly reminiscent of measurement tasks found throughout the social sciences. With this in mind, we present a framework, grounded in measurement theory from the social sciences, for measuring concepts related to the capabilities, impacts, opportunities, and risks of GenAI systems. The framework distinguishes between four levels: the background concept, the systematized concept, the measurement instrument(s), and the instance-level measurements themselves. This four-level approach differs from the way measurement is typically done in ML, where researchers and practitioners appear to jump straight from background concepts to measurement instruments, with little to no explicit systematization in between. As well as surfacing assumptions, thereby making it easier to understand exactly what the resulting measurements do and do not mean, this framework has two important implications for evaluating evaluations: First, it can enable stakeholders from different worlds to participate in conceptual debates, broadening the expertise involved in evaluating GenAI systems. Second, it brings rigor to operational debates by offering a set of lenses for interrogating~the~validity of measurement instruments and their resulting measurements.\looseness=-1
\end{abstract}

\section{Measurement and its Role in Evaluating GenAI Systems}

Evaluating an ML system means making evaluative judgements about that system's capabilities, impacts, opportunities, and risks 
in order to facilitate decisions like whether it should be used for a particular purpose, whether it should be deployed in a particular context, or even whether it should be redesigned.
However, we cannot make such evaluative judgements without accurate information about systems' capabilities, impacts, opportunities, and risks. 
Often, this information takes the form of \emph{measurements} on nominal, ordinal, interval, and ratio scales, where each measurement reflects the amount of some \emph{concept} of interest---be it a concept related to capabilities, like reasoning skills; a concept related to impacts, like causing a user to feel harmed; a concept related to opportunities, like the possibility of helping a user complete a certain type of task; or a concept related to risks, like the possibility of privacy violations. Such measurements are obtained via the \emph{process of measurement}, which can involve both qualitative and quantitative approaches. Thus, measurement is often central to evaluation. \looseness=-1

Across academia, industry, and government~\citep[e.g.,][]{nistairmf, cooper2023genlaw, perez2022redteaminglanguagemodels}, there is an increasing awareness that the measurement tasks involved in evaluating generative AI (GenAI) systems are especially difficult---more so than those involved in evaluating supervised ML systems. This is because the concepts to be measured tend to be complex and nuanced, and may even have contested meanings~\citep[e.g.,][]{mulliganfairness, mulligan2016privacy} across and within use cases, cultures, and languages. Although ML researchers and practitioners have proposed myriad approaches and instruments intended to measure such concepts,
it is often very difficult to know whether these approaches and instruments yield reliable and valid measurements.\looseness=-1

We argue that the 
measurement tasks involved in evaluating GenAI systems are highly reminiscent of measurement tasks found throughout the social sciences. Social scientists have been thoughtfully measuring complex and contested concepts---ideology, democracy, media bias, framing, to name a few---for over fifty years~\citep[e.g.,][]{berelson1952content, zaller1992nature}. Thus, our perspective is that 
the ML community would benefit from learning from and drawing on the social sciences when developing approaches and instruments for measuring concepts related to the capabilities, impacts, opportunities, and risks of GenAI systems.\looseness=-1

\section{A Measurement Framework for GenAI Systems}
\label{sec:framework}

When measuring complex and contested concepts, social scientists often turn to \emph{measurement theory}, which offers a framework for articulating distinctions between concepts and their operationalizations via \emph{measurement instruments}---i.e., the procedures and artifacts used to obtain measurements of those concepts, such as classifiers, annotation guidelines, scoring rules---and a set of lenses for interrogating~the validity of measurement instruments and their resulting measurements~\citep[e.g.,][]{adcock2001measurement,cronbach1955construct,messick1996validity}.\looseness=-1

\begin{figure}[h]
    \centering    \includegraphics[width=0.75\linewidth]{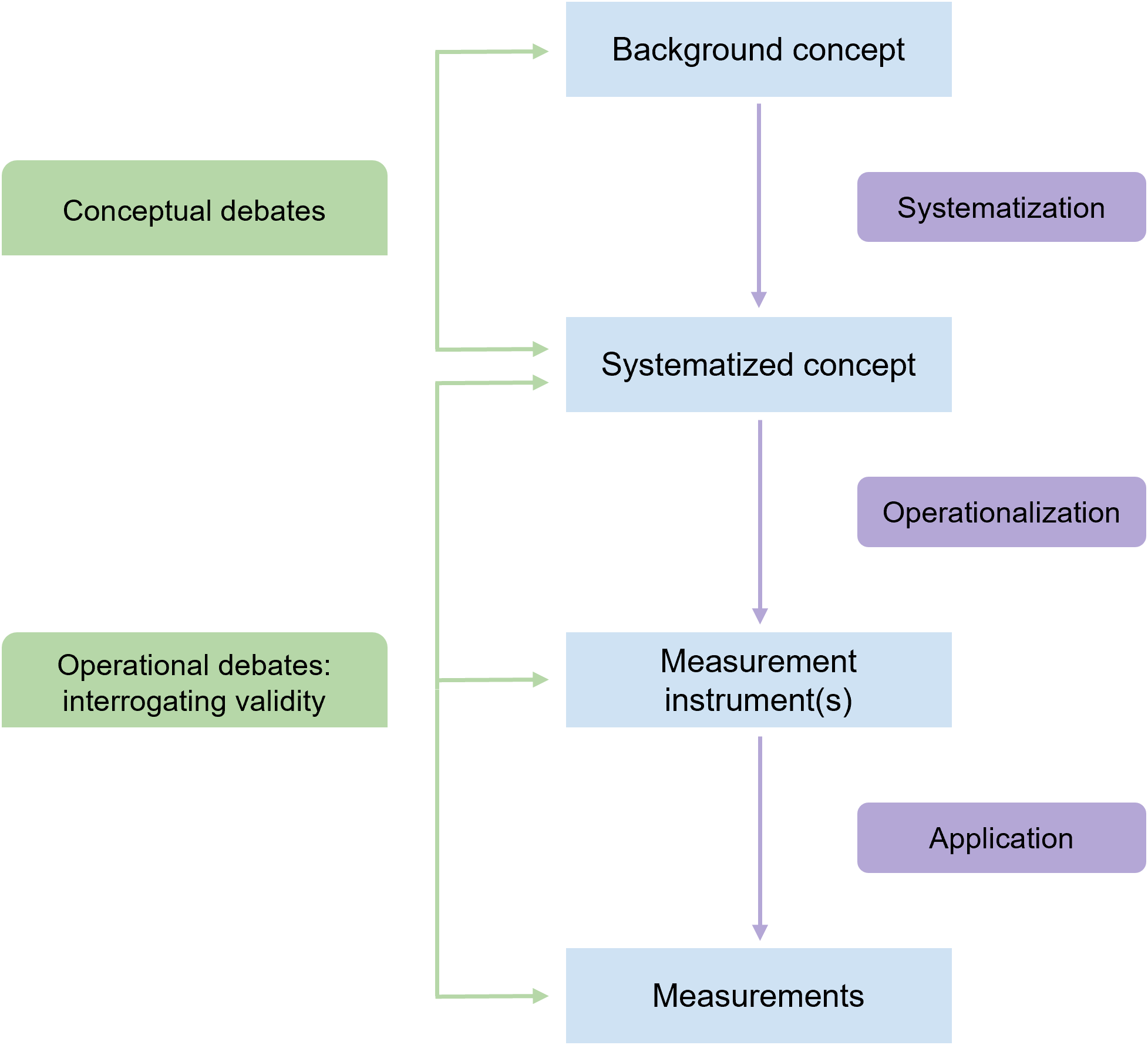}
    \vspace{0.25cm}
    \caption{Measurement in the social sciences, as formulated by \citet{adcock2001measurement}. The four levels---the background concept, the systematized concept, the measurement instrument(s), and the instance-level measurements---are linked by three processes---systematization, operationalization, and application. We also indicate where conceptual debates and operational debates occur. 
    Note that we use slightly different terminology to Adcock and Collier; the ideas remain unchanged, however.\looseness=-1}
    \label{measurement-schematic}
\end{figure}

The framework, as formulated by \citet{adcock2001measurement}, provides a structured approach for 
producing measurements 
that reflect complex concepts. It distinguishes between four levels: the \emph{background concept} or ``broad constellation of meanings and understandings associated with [the] concept;'' the \emph{systematized concept} or ``specific formulation of the concept[, which] commonly involves an explicit definition;'' the \emph{measurement instrument(s)} used to produce instance-level measurements; and the \emph{instance-level measurements} themselves~\citep{adcock2001measurement}. These four levels are linked by three processes: \emph{systematization}, \emph{operationalization}, and \emph{application}, as shown in Figure~\ref{measurement-schematic}.
Note that we use slightly different terminology to Adcock and Collier; the ideas remain unchanged, however.
For example, when measuring the prevalence of demeaning text generated by a GenAI system, the background concept encompasses all possible definitions of ``demeaning text.'' From here, we might select a single definition like ``system [...] outputs with dehumanizing or offensive associations, or which otherwise threaten people's sense of security or dignity''~\citep{blodgett2021diss}. However, since this definition itself encompasses a broad range of meanings and understandings, it must be further systematized, perhaps into a set of 
linguistic patterns that equate a particular social group to an animal, advocate for animal-like treatment of the group, equate the group to an inanimate object, note qualities of the group that are like those of an inanimate object, equate the group to a disease or disorder, etc.~\citep{corvi2024}. Collectively, these linguistic patterns constitute the systematized concept. Finally, we might operationalize this systematized concept via an ML classifier trained to identify each of these linguistic patterns in system outputs, resulting in instance-level measurements that comprise a set of binary labels indicating the presence or absence of one or more of the linguistic  patterns in each system output.\looseness=-1

This approach differs from the way measurement is typically done in ML,
where researchers and practitioners appear to jump straight from background concepts to measurement instruments, with little to no explicit systematization in between~\citep[e.g.,][]{blilihamelin2023intelligence, cooper2021emergent, jacobs2021measurement, blodgett2020language,EBCD}. However, the systematization process is particularly 
important when measuring complex and contested concepts like those related to the capabilities, impacts, opportunities,  and risks of GenAI systems. Without an explicitly systematized concept, it is hard to know exactly 
what is being operationalized, and thus measured. For example, StereoSet~\citep{nadeem2020stereoset} and CrowS-Pairs~\citep{nangia2020crows}, two widely used benchmarks in NLP for measuring stereotyping, appear to jump straight from high-level definitions of the concept, encompassing broad constellations of meanings and understandings, to specific measurement instruments, obscuring exactly what those instruments  measure~\citep{blodgett2021stereotyping}. 
Both benchmarks' measurement instruments rely on crowdworkers, who, in the absence of an explicitly 
systematized concept, must rely on their own understandings of these high-level definitions, which may be contradictory.
A similar critique also applies to recent work on measuring stereotypes in the context of text-to-image generation~\citep{jha2024visageglobalscaleanalysisvisual, cho2023dallevalprobingreasoningskills}.\looseness=-1  

\section{Evaluating Evaluations of GenAI Systems}

This framework surfaces assumptions, thereby making it easier to understand exactly what the resulting measurements do and do not mean. It also separates conceptual debates---i.e., does our systematized concept capture what we want it to capture?---from operational debates---i.e., did we operationalize the systematized concept in a way that yields reliable and valid measurements? We argue that this separation has two important implications for evaluating evaluations, outlined below.\looseness=-1

First, systematization can enable stakeholders from different worlds---e.g., open-source developers, policymakers, 
users, 
members of marginalized communities, all of whom may be interested in measuring a concept for different reasons---to participate in conceptual debates and thus advocate for particular meanings and understandings. Measuring a complex and contested concept necessarily means making choices
about which of its meanings and understandings will be reflected in the resulting measurements and which will not. For example, hate speech can be understood from a practical perspective as speech that promotes violence~\citep{youtube} or from a normative perspective as
speech that ``den[ies] the basic standing of [...] vulnerable social groups''~\citep{lepoutre}.
Without an explicitly systematized concept, many of these choices are accessible only indirectly via the measurement instrument(s), which may be hard for stakeholders other than ML researchers and practitioners to engage with. We therefore argue that systematization can help broaden the expertise involved in 
evaluating 
GenAI systems.\looseness=-1

Second, systematization brings rigor to operational debates. When measuring complex and contested concepts, there are no directly observable, universally agreed-upon labels or scores against which to evaluate the resulting measurements, making operational debates fraught. Measurement theory therefore 
offers a set of lenses for interrogating the validity of measurement instruments and their resulting measurements: \emph{face validity}, \emph{content validity}, \emph{convergent validity}, \emph{discriminant validity}, \emph{predictive validity}, \emph{hypothesis validity}, and \emph{consequential validity}~\citep[e.g.,][]{jacobs2021measurement}. Each lens constitutes
a different source of evidence about validity. For example, content validity focuses on whether a measurement instrument captures all relevant aspects of a systematized concept, while convergent validity focuses on whether the resulting measurements are similar to measurements obtained using other (already validated) instruments for measuring that systematized concept.
Distinguishing between the background concept and the systematized concept is  crucial to obtaining meaningful evidence about validity using this set of lenses: ``If [we] seek to establish [...] validity in relation to a background concept with multiple competing meanings, [we] may find a different answer [...] for each meaning''~\cite{adcock2001measurement}. As a result, we again argue that  systematization is important,
this time because the existence of an explicitly systematized concept enables these lenses to play a crucial role in evaluating evaluations of GenAI systems.\looseness=-1

\section{Broader Impacts and Limitations}

In calling on the ML community to learn from and draw on the social sciences when developing approaches and instruments for measuring concepts related to the capabilities, impacts, opportunities, and risks of GenAI systems, we may be misunderstood as recommending   
that the ML community 
adopt existing measurement instruments from the social sciences. 
Rather, we advocate for adopting the \emph{framework} that social scientists often turn to for measurement; we do not advocate for na\"{i}vely transferring measurement instruments designed for humans (e.g., competency tests) to the context of GenAI systems. Effectively adapting existing measurement instruments requires carefully thinking through precisely the kinds of conceptual and operational questions that the framework described in Section~\ref{sec:framework}  highlights. In this regard, our perspective is similar to that of \citet{wang23evaluating}, who advocate for taking a construct-oriented approach when evaluating GenAI systems by drawing on psychometrics; they too caution against na\"{i}vely using measurement instruments designed for humans.\looseness=-1

Likewise, in suggesting that the framework described in Section~\ref{sec:framework} can make evaluations of GenAI systems more rigorous, we do not mean to suggest that better measurements will inevitably improve how GenAI systems are developed, deployed, used, or regulated. The social sciences themselves have repeatedly demonstrated that better understanding of a problem does not automatically translate into better policy. Although using the framework can help clear up conceptual confusion, broaden the expertise involved in evaluating GenAI systems, and yield more valid measurements,  it needs to be accompanied by sustained efforts to meaningfully inject research into policymaking and practice.

Although the measurement approaches and instruments proposed by ML researchers and practitioners tend to be quantitative, the process of measurement can involve both qualitative and quantitative approaches. As  result, we emphasize that the framework described in Section~\ref{sec:framework} supports both qualitative and quantitative approaches. Indeed,  
\citet{adcock2001measurement} stated that 
their framework, which forms the basis of ours, was intended to be a shared standard that would allow ``quantitative and qualitative scholars to assess more effectively, and communicate about, issues of valid measurement.''

Finally, we stress that our suggestions are not a panacea. Even when evaluations of GenAI systems are grounded in measurement theory, they may fall short of what we would like them to accomplish. If anything, the framework described in Section~\ref{sec:framework} will often reveal the shortcomings of evaluations---i.e., the ways they depart from what their designers hoped to achieve. Rather than thinking of measurement theory as a solution to all the problems that beset evaluations of GenAI systems, we think of it as~a~way~to appropriately and precisely qualify exactly what measurement instruments measure.\looseness=-1

\section*{Acknowledgments}

This work was supported in part by the 
Microsoft Research AI \& Society fellows
program.

\bibliographystyle{plainnat}
\bibliography{refs,measmtrunningcombinedbib}

\end{document}